# Crystal structure and phase transitions at high pressures in the superconductor FeSe$_{0.89}$S$_{0.11}$

Yulia A. Nikiforova[1], Anna G. Ivanova[1], Kirill V. Frolov[1], Igor S. Lyubutin[1], Dmitriy A. Chareev[2], Arseniy O. Baskakov[1], Sergey S. Starchikov[1], Ivan A. Troyan[1], Mariana V. Lyubutina[1], Pavel G. Naumov[1], and Mahmoud Abdel-Hafiez[3]

[1] *Shubnikov Institute of Crystallography of FSRC "Crystallography and Photonics" of Russian Academy of Sciences, 119333 Moscow, Russia*
[2] *Institute of Experimental Mineralogy of Russian Academy of Sciences, 123456 Chernogolovka, Moscow District, Russia*
[3] *Department of Physics and Astronomy, Box 516, Uppsala University, Uppsala, SE-75120, Sweden.*

**Abstract**

We report on the structural phase transitions in the FeSe$_{0.89}$S$_{0.11}$ superconductor with $T_C$ = 11 K by powder synchrotron X-ray diffraction at high pressures up to 18.5 GPa under compression and decompression modes. In order to create high quasi-hydrostatic pressures, diamond anvil cells (DACs) filled with helium as a pressure transmitting medium were used. It was found that at ambient pressure and room temperature, FeSe$_{0.89}$S$_{0.11}$ has a tetragonal structure (space group *P4/n*). Under compression, in the region of 10 GPa, a phase transition from the tetragonal into the orthorhombic structure (sp. gr. *Pnma*) is observed, which persists up to 18.5 GPa. Our results strongly suggest that, at decompression, as the applied pressure decreases to 6 GPa and then is completely removed, most of the sample recrystallizes into the hexagonal phase of the structural type NiAs (sp. gr. *P6$_3$/mmc*). However, the other part of the sample remains in the high pressure orthorhombic phase (sp. gr. *Pnma*), while the tetragonal phase (sp. gr. *P4/n*) is not restored. These observations illustrate a strong hysteresis of the structural properties of FeSe$_{0.89}$S$_{0.11}$ during a phase transition under pressure.

## 1. Introduction

Iron chalcogenides have been intensively studied since the discovery of superconductivity in FeSe$_{1-x}$ in 2009 [1]. At ambient pressure and room temperature, FeSe has a tetragonal structure, the layers of which, consisting of FeSe$_4$ tetrahedra joined by edges, are interconnected by weak Van der Waals forces. The physical and chemical properties of superconductors based on transition metal chalcogenides and, in particular, iron-containing superconductors are very sensitive to stoichiometry, crystal structure defects, and doping [2–4]. Superconductivity in FeSe$_{1-x}$ is achieved in a narrow range of stoichiometry. For example, in FeSe$_{0.99}$, the critical temperature of the superconducting transition $T_C$ is 8.5 K, while in the FeSe$_{0.97}$ compound there is no superconducting transition down to 2 K [2]. Such an effect of stoichiometry on superconductivity may be due to the crystal structure of these compounds. FeSe$_{0.99}$ undergoes a structural phase transition at ~ 90 K from the tetragonal modification of α-FeSe (sp. gr. *P4/nmm* [1,5]) to the orthorhombic phase of β-FeSe (sp. gr. *Cmma*) [2], while FeSe$_{0.97}$ remains in the initial tetragonal phase down to 20 K.

Along with this, external pressure significantly affects the superconducting properties of FeSe-based compounds. Thus, an increase in $T_C$ from 8.5 K (at ambient pressure) to 36.7 K at a pressure of 8.9 GPa was observed in [6]. However, a further increase in pressure gradually decreases the $T_C$ value, and at 29 GPa, superconducting transition is already absent. In iron selenide at a pressure of 12 GPa, a new δ-phase of the NiAs-type with hexagonal symmetry (sp. gr. *P6$_3$/mmc*) appears, which coexists with the initial α- or β- phase (depending on temperature) with a further increase in pressure to 38 GPa [6]. Meanwhile, hexagonal FeSe crystals can exist at ambient pressure and temperature [7]. Simultaneously, in the FeSe compound, the phase transition from the tetragonal phase to the orthorhombic one of the MnP-type (sp. gr. *Pnma*) occurs at pressures of 6.7–8.1 GPa [8,9]. The superconducting properties of FeSe can also be affected by the so-called chemical pressure, i.e. by doping with elements with different ionic radii. Thus, substitution of Se-atoms with Te-atoms increases $T_C$ from 8.5 to 13 K [10]. Since the ionic radius of the Te is larger than Se, on can speak of a "negative" chemical pressure created by Te-atoms inside the crystalline structure. Substitution of Se-atoms with S-atoms also increases $T_C$ and the upper critical magnetic field $H_{C2}$ [11]. The ionic radius S is smaller than Se; therefore, it is assumed that S-atoms create a "positive" chemical pressure. The partial replacement of iron cations in the FeSe structure with transition metal cations can also significantly affect the superconducting properties of iron selenide. For example, in the $Cu_xFe_{1.01-x}Se$ compound, the substitution of Fe by Cu leads to a metal-insulator transition even at low concentration $x = 0.03$, and superconductivity is completely suppressed [12]. However, an applied external pressure of 1.5 GPa recovers superconductivity in this crystal with $T_C = 6.6$ K [13].

Therefore, the combination of external and chemical pressures is of great interest for studying the mechanisms of superconductivity in iron-containing superconductors. In Refs.[14, 15], pressure – temperature *(P-T)* phase diagrams of $FeSe_{1-x}S_x$ were obtained through measurements of transport properties at high pressures. It turned out that the combination of external and chemical pressures can affect the position of structural and superconducting phases in *P-T* diagrams and vary the value of $T_C$. It should be noted the fragmentation of X-ray diffraction studies of FeSe single crystals carried out to date. This leads to different interpretations of the *P-T* phase diagram of FeSe. Thus, the structure of single-crystal iron selenide $FeSe_{0.96}$ was studied by powder X-ray diffraction in [5]. The structural phase diagram *P(T)* of undoped FeSe studied using single-crystal diffraction was presented in [8]. The data on single crystal diffraction of FeSe1-xSx are presented in [14] only for pressures up to 0.8 GPa. Therefore, it is a matter of interest to see how the crystal structure will evolve of $FeSe_{1-x}S_x$ up to high pressures. Pressure has long been recognized as a fundamental thermodynamic variable, which is a convenient tool for deep understanding of various SC characteristics. It is considered as a clean way to tune basic electronic and structural properties without changing the stoichiometry.

In the present work, a structural study of $FeSe_{0.89}S_{0.11}$ by single crystal X-ray diffraction at ambient and high pressures was performed. Also, phase transitions in $FeSe_{0.89}S_{0.11}$ under pressures up to 18.5 GPa were studied by X-ray powder diffraction using synchrotron radiation in compression and decompression modes. At ambient pressure / room temperature the investigated system, $FeSe_{0.89}S_{0.11}$, presents a tetragonal structure, while at high pressure, we observed a phase transition from the tetragonal into the orthorhombic structure and the phase transition persists up to 18.5 GPa.

## 2. Experimental details

Layered single crystals of iron selenide with partial replacement of selenium by sulfur Fe $Fe(Se_{0.89\pm0.01}S_{0.11\pm0.01})_{1-\delta}$ were grown in the form of plates in pumped quartz ampoules using $AlCl_3$ / KCl flux. The temperature gradient when the hot and cold ends of the ampoule were at 400 °C and 350 °C, respectively, was maintained for 45 days. Details of the technique are described in [16]. The chemical composition of the crystals and the uniformity of the sulfur distribution in the sample were studied in detail in [11] using a TESCAN Vega II XMU scanning electron microscope. It was found that sulfur is evenly distributed (within the error range) and the sample has the composition $Fe(Se_{0.89\pm0.01}S_{0.11\pm0.01})_{1-\delta}$. Subsequently, when analyzing the results of structural measurements under pressure, we were based on the indicated composition.

The heat capacity and resistivity were measured in the temperature range of 2–300 K using the Quantum Design Physical Property Measurement System (PPMS). Resistance was measured by the standard four-probe method with direct current (D.C.) applied in the (*ab*) plane of crystals, similarly to [17]. X-ray diffraction data of one of the $FeSe_{1-x}S_x$ single crystals with the maximum sulfur content ($FeSe_{0.89}S_{0.11}$) were collected at the SNBL ID-01 station of the European source of synchrotron radiation ESRF (beamline ID-01) using a PILATUS@SNBL diffractometer ($\lambda = 0.7458$ Å) at ambient pressure and room temperature. Processing of the single-crystal x-ray diffraction data with the use of emperical absorption correction was performed using the CrysAlis Pro program [18]. The crystal structure was determined by the direct method and refined using the ShelxS and ShelxL programs, respectively [19] as part of the Olex2 software package [20].

The study of $FeSe_{0.89}S_{0.11}$ crystal structure at high pressures up to 18.5 GPa was carried out under compression and decompression conditions in a high-pressure membrane diamond anvil cell (DAC) having a working platform (culet) diameter of 300 μm. Rhenium gasket with a central hole of 176 μm was used. The pressure was measured by the shift of the luminescence band of a ruby placed in the working volume of the cell next to the sample [21,22]. Helium was used as the pressure transmitting medium, which ensured hydrostatic compression of the sample. The initial pressure was set to 0.15 GPa. Powder X-ray diffraction patterns in the DAC were obtained at the ID-27 synchrotron station in ESRF (Grenoble, France) using a focused (2 × 3 μm$^2$) monochromatic beam (33 keV, λ = 0.3738 Å) and a two-dimensional MAR CCD 165 detector located at a distance 215.7 mm from the sample. The exposure times was 120 seconds. Analysis and integration of powder x-ray diffraction data were performed in the Dioptas 5.0 program [23]. The unit cell parameters were determined from powder X-ray diffraction patterns using the DICVOL14 [24] and PreDICT [25] programs. The refinement of the unit cell parameters of the high-pressure phases was carried out as a result of a full-profile analysis by the Le Bail method [26] in the JANA2006 program [27].

## 3. Results

*3.1 Heat capacity and electrical resistance*

According to the results of calorimetric measurements, at ambient pressure in FeSe$_{0.89}$S$_{0.11}$, several phase transitions are observed with decreasing temperature. Figure 1 shows low-temperature specific heat measured at zero-filed of FeSe and FeSe$_{0.89}$S$_{0.11}$ single crystals. In zero-field specific-heat measurements, a very sharp anomaly is clearly seen of both samples. This anomaly is attributed to the superconducting transition at T$_c$. The specific heat jump confirms bulk superconductivity in the investigated systems. The estimated universal parameter $\Delta C/\gamma_n T_c$ of the specific heat at $T$c is ≈ 2.14 and 1.95 mJ/mol K$^2$ for $x$=0 and 0.11, respectively. These values are higher than the prediction of the weak-coupling Bardeen-Cooper-Schrieer (BCS) theory ($\Delta C$el/ $\gamma_n T_c$ = 1.43).

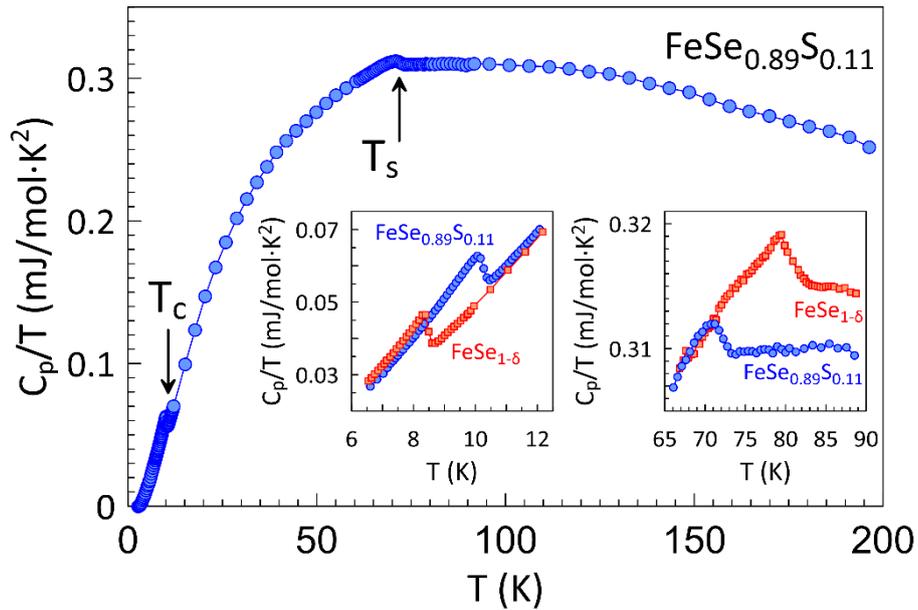

Fig. 1. Temperature dependence of specific heat at ambient pressure for FeSe$_{0.89}$S$_{0.11}$. $T_S$ is the temperature of the structural transition, $T_C$ is the temperature of the superconducting transition. For comparison, the insets show the dependences $C_p$(T) for FeSe$_{1-\delta}$ and FeSe$_{0.89}$S$_{0.11}$ in the temperature range of the transition to the superconducting state and during the structural transition from the tetragonal phase to the orthorhombic one

As was shown in Ref. [2], in the FeSe crystal, the transition in the region $T_S$ ≈ 80 K is due to the transformation of the tetragonal structure into the orthorhombic one. The nature of the heat capacity anomaly detected in our FeSe$_{0.89}$S$_{0.11}$ sample at $T_S$ ≈ 71 K (Fig. 1) is similar to that observed in FeSe. This allows concluding that the FeSe$_{0.89}$S$_{0.11}$ sample undergoes a structural phase transition to the orthorhombic phase at a lower temperature $T_S$ ≈ 71 K [11,28]. An anomaly in the dependence of electrical resistance is also observed in this temperature range (Fig. 2). Meanwhile, the superconducting transition shifts toward higher temperatures with increasing sulfur concentration, and for FeSe$_{0.89}$S$_{0.11}$, the $T_C$ value is 11 K (Fig. 2) [11, 17]. The application of external pressure can induce a change in the crystal structure and significantly affect the superconducting properties [9, 11].

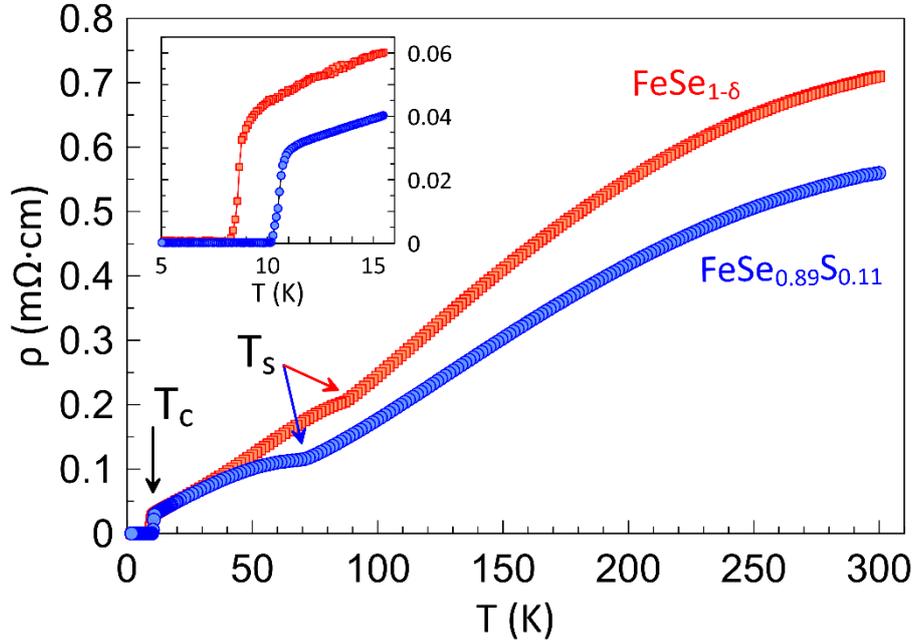

Fig. 2. Temperature dependence of resistivity for the FeSe$_{1-\delta}$ and FeSe$_{0.89}$S$_{0.11}$ samples in zero filed. $T_S$ is the temperature of the structural transition, $T_C$ is the temperature of the superconducting transition. The inset on an enlarged scale shows the $\rho$(T) dependences for FeSe$_{1-\delta}$ and FeSe$_{0.89}$S$_{0.11}$ in the temperature range of the transition to the superconducting state.

*3.2 Structural analysis data*

A single-crystal X-ray diffraction measurement of the FeSe$_{0.89}$S$_{0.11}$ sample was carried out at ambient pressure. The parameters of the tetragonal unit cell were determined from 376 reflections: $a$ = 3.8093(4) Å and $c$ = 5.529 (1) Å. From the analysis of systematic extinction of reflections, the possible space groups *P*4*/nmm* and *P*4*/n* were determined. A three-dimensional set of intensities was integrated and averaged over equivalent reflections in both space groups. The resulting *R*-factors, when averaging equivalent reflections, were 15% and 3.5% in sp.gr. *P*4*/nmm* and *P*4*/n*, respectively. Therefore, the structure was identified in the space group *P*4*/n* by (over) 100 independent reflections with $I > 3\sigma$, where $I$ is the intensity of the reflection, $\sigma$ is the standard deviation. The atomic coordinates of Fe (2*b*: 1/4, 3/4, 1/2) and Se (2*c*: 1/4, 3/4, 0.2333 (2)) were determined by direct methods in the ShelxS program [19]. The coordinates of the atoms and the anisotropic parameters of the atomic displacement were refined using the ShelxS program [19] to the *R* factor of 3.9%. The composition of FeSe$_{0.89}$S$_{0.11}$ was confirmed by the refinement of the occupancies of both atoms S and Se in the same wyckoff position 2*c*. The crystal structure of FeSe$_{0.89}$S$_{0.11}$ of the tetragonal modification *P*4*/n* is shown in Fig. 3.

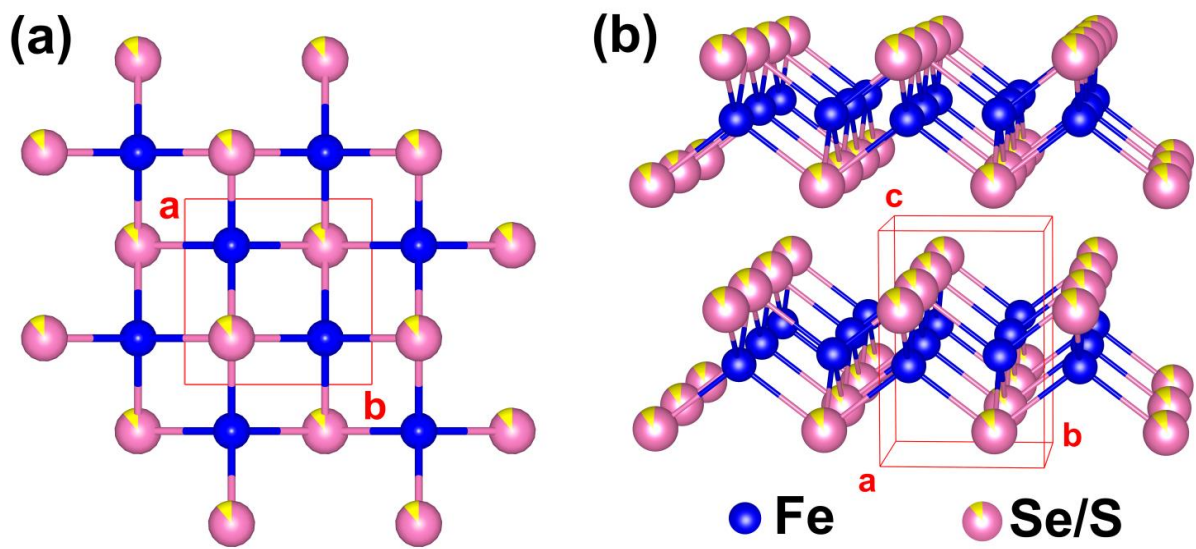

Fig. 3. The crystal structure of the tetragonal modification of FeSe$_{0.89}$S$_{0.11}$ (sp. gr. *P*4/*n*): (a) - in the projection onto the (110) plane; (b) - view in the [100] direction.

The evolution of X-ray diffraction patterns of a polycrystalline FeSe$_{0.89}$S$_{0.11}$ sample under pressure is shown in Fig. 4. The two-dimensional pattern and the refined parameters of the tetragonal phase *P*4/*n* at 0.15 GPa ($a = 3.7637(4)$ Å и $c = 5.4862(3)$ Å) are shown in Fig. 5a and 5b, respectively. It was found that with an increase in pressure up to 10.5 GPa, the initial tetragonal modification FeSe$_{0.89}$S$_{0.11}$ (sp. gr. *P*4/*n*) remains stable. At a pressure of 10.5 GPa, the X-ray diffraction completely changes (Fig. 4); all the reflections of the tetragonal phase disappear and new reflections appear, which indicates a phase transition with a rearrangement of the crystal structure.

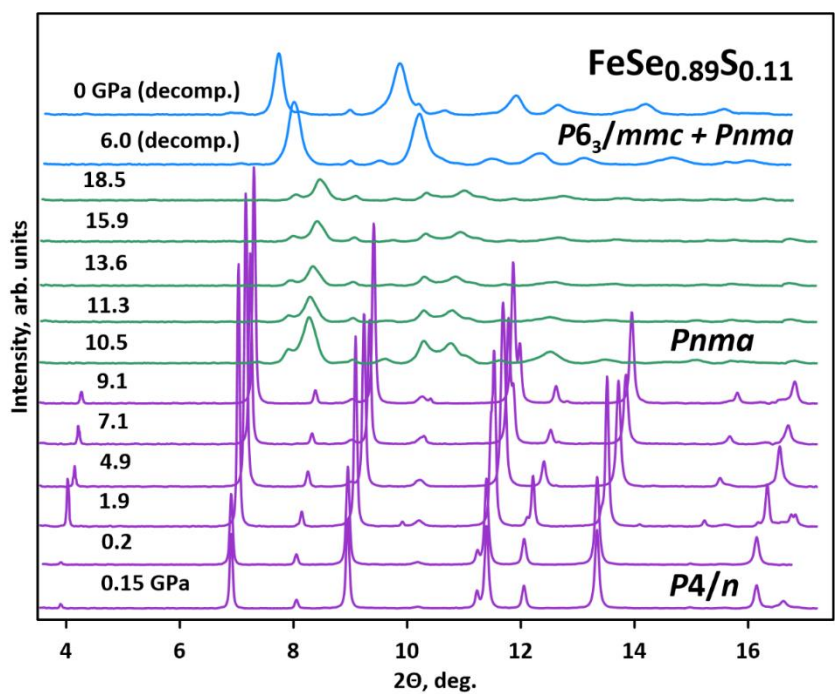

Fig. 4. Evolution of X-ray diffraction patterns of FeSe$_{0.89}$S$_{0.11}$ under increasing pressure up to 18.5 GPa and under decompression.

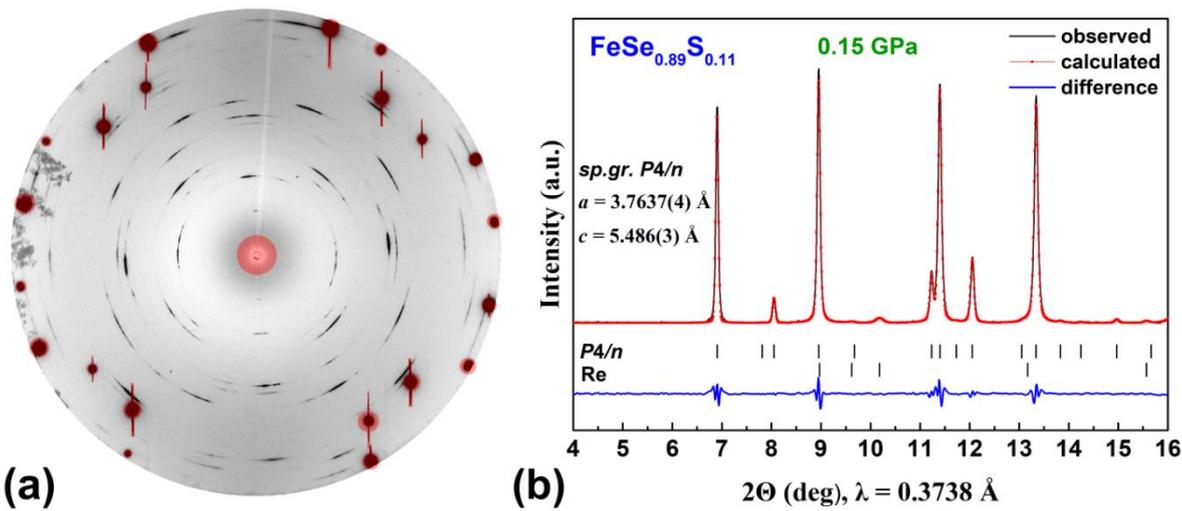

Fig. 5. Two-dimensional X-ray diffraction pattern of FeSe$_{0.89}$S$_{0.11}$ at a pressure of 0.15 GPa (a). Red spots show a mask superimposed on peaks (reflections) from diamond anvils during integration. (b) Full-profile refinement of the cell parameters of the tetragonal modification in the *P*4/*n* space group by the Le Bale method.

Figure 6a shows a two-dimensional diffraction pattern of FeSe$_{0.89}$S$_{0.11}$ at 11.3 GPa, which is characteristic of the entire range of applied pressures above 10.5 GPa. Reflections in patterns measured in the pressure range 10.5 - 18.5 GPa are satisfactorily displayed in the parameters of the orthorhombic structure with sp. gr. *Pnma* (structural type of MnP). A similar structure was found in [8] for FeSe at pressures of 6.7–8.1 GPa. Fig. 6b shows the result of a full-profile refinement of the parameters of the orthorhombic phase cell at 11.3 GPa.

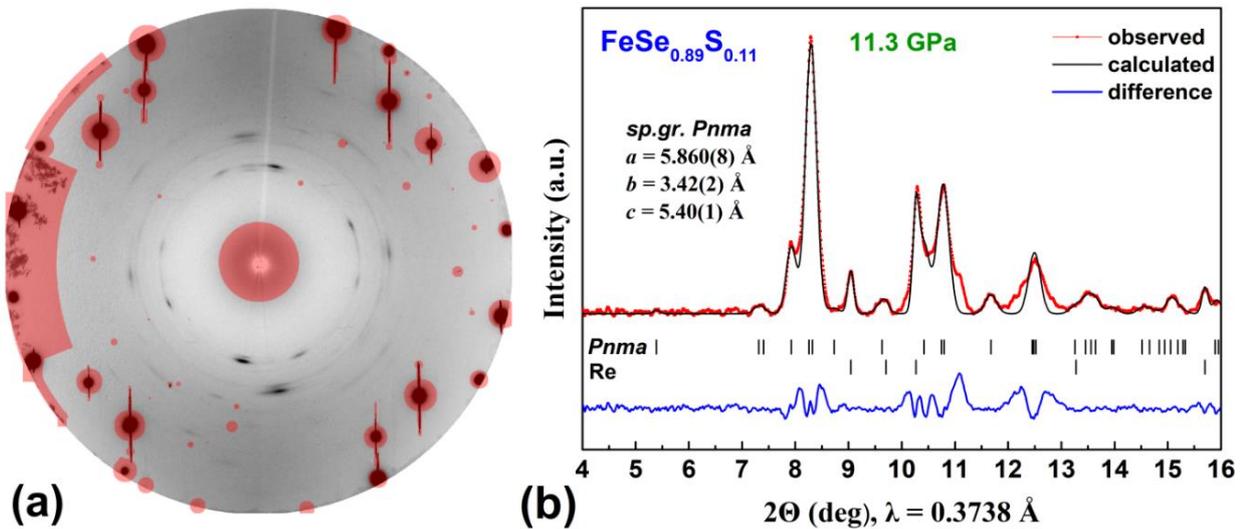

Fig. 6. Two-dimensional X-ray diffraction pattern of the FeSe$_{0.89}$S$_{0.11}$ crystal at a pressure of 11.3 GPa (a). Red spots show a mask superimposed on peaks (reflections) from diamond anvils when integration. (b) Full-profile refinement of the parameters of the orthorhombic modification cell by the Le Bale method in the *Pnma* space group. Unit cell parameters: *a* = 5.860(8) Å, *b* = 3.42(2) Å, *c* = 5.40(1) Å, *Z* = 4. The apparent broadening of the reflections and a decrease in their intensities are noticeable in comparison with those of the tetragonal structure at pressures below 10 GPa shown in Figs. 4 and 5.

The crystal structure of the orthorhombic modification FeSe$_{0.89}$S$_{0.11}$ at pressures above the phase transition *P* > 10 GPa is shown in Fig. 7. It can be seen from the projection onto the (011) plane (Fig. 7a) that the orthorhombic structure of FeSe$_{0.89}$S$_{0.11}$ has a pronounced pseudo-hexagonal motif, since the periodicity of the arrangement of Fe and Se/S atoms in the structure can be represented by a distorted hexagonal lattice. The transition from the tetragonal (see Fig. 3) to the orthorhombic modification is accompanied by an increase in the coordination number of Fe (from 4 to 6). In the structure of the orthorhombic modification FeSe$_{0.89}$S$_{0.11}$, Se/S atoms form a distorted hexagonal close-packed structure, where Fe atoms are located in octahedral sites (hollows) (Fig. 7b).

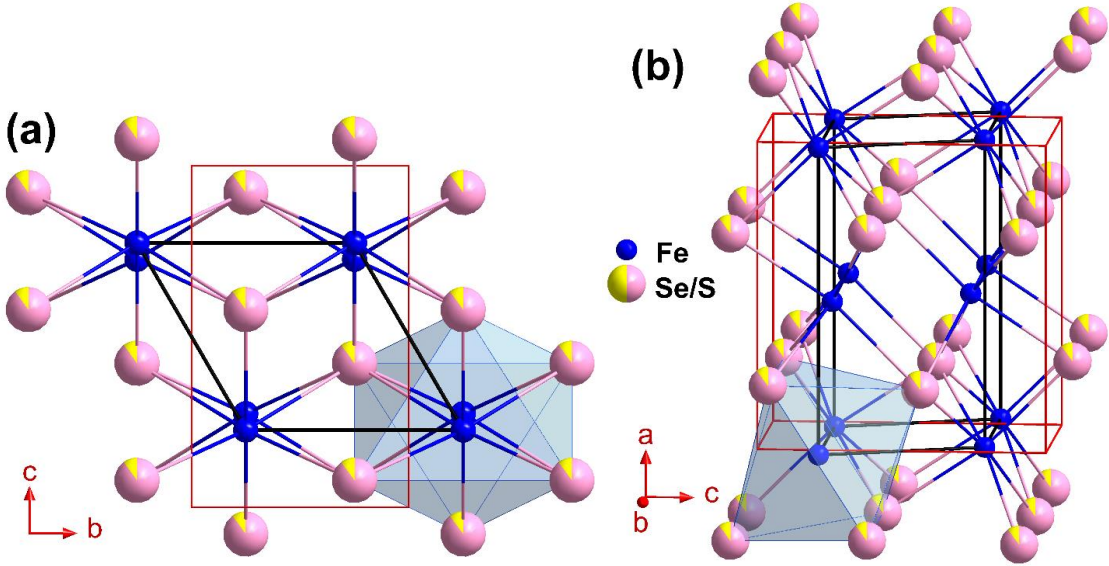

Fig. 7. The crystal structure of the orthorhombic modification FeSe$_{0.89}$S$_{0.11}$ (sp. gr. *Pnma*) at pressures above the phase transition $P > 10$ GPa : (a) is in the projection onto the (011) plane, (b) is in the isometric projection onto the (110) plane. The red outline shows an orthorhombic unit cell. The black outline shows a distorted hexagonal unit cell.

After a phase transition above 10 GPa, a gradual broadening of the reflections and a noticeable decrease in their intensity are observed in FeSe$_{0.89}$S$_{0.11}$ diffraction patterns, which indicates partial amorphization of the sample under pressure. A similar change in the diffraction patterns with increasing pressure was previously described for iron selenide [29,30].

Upon decompression of the FeSe$_{0.89}$S$_{0.11}$ sample, X-ray diffraction patterns were obtained at applied pressures of 6 GPa and 0 GPa. Figure 8 shows a two-dimensional X-ray pattern and the result of a full-profile refinement of the unit cells parameters of the FeSe$_{0.89}$S$_{0.11}$ phases at 6 GPa. We found that during decompression, the intensity of reflections increases (Figs. 8a and 4), which indicates an increase in the degree of crystallinity of the sample. All intense reflections are indexed in a hexagonal cell with sp. gr. *P6$_3$/mmc*, but there are also residual less intense reflections from the orthorhombic phase (Fig. 8b) characteristic of the pressure region above the structural transition. Changes observed under decompression in the diffraction patterns of FeSe0.89S0.11 can be explained by recrystallization of the sample, when the crystal structure is transformed from orthorhombic to hexagonal with a NiAs type structure. The hexagonal phase is retained when the external pressure is completely relieved, and its unit cell parameters at $P = 0$ are $a = 3.603(6)$ Å and $c = 6.073(9)$ Å. Reflections from the orthorhombic phase also persist with a decrease in pressure to ambient, which indicates a large hysteresis of structure parameters.

Remarkably, that in the diffractogram there are no reflections from the initial tetragonal phase FeSe$_{0.89}$S$_{0.11}$ (Fig. 4). We note that the appearance of a hexagonal modification of FeSe with a NiAs type structure was previously observed and described in Ref. [6] at a pressure of 12 GPa and in Ref. [31] at a pressure of 10 ± 2 GPa.

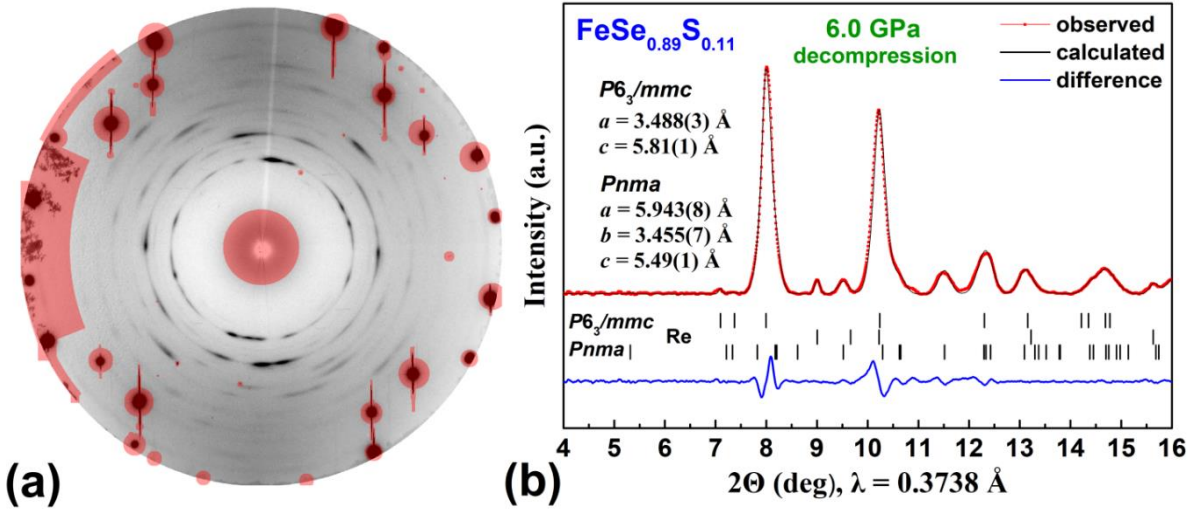

Fig. 8. Two-dimensional X-ray diffraction pattern of the FeSe$_{0.89}$S$_{0.11}$ crystal at a pressure of 6.0 GPa in the decompression mode (a) and full-profile refinement of the hexagonal modification by the Le Bale method (b). The parameters of the hexagonal phase with the structural type NiAs are $a$ = 3.488(3) Å, $c$ = 5.81(1) Å, $Z$ = 2. The unit cell parameters of the orthorhombic phase $Pnma$ are $a$ = 5.943(8) Å, $b$ = 3.455(7) Å, $c$ = 5.49(1), $Z$ = 4.

*3.3 Equation of state*

According to the results of structural analysis of the FeSe$_{0.89}$S$_{0.11}$ crystal, the dependence of the unit cell volume on pressure $V(P)$ (equation of state) at room temperature was constructed (Fig. 9). When approximating the dependence $V(P)$ in the pressure range 0 - 10 GPa, the equation of state in the form of Birch-Murnaghan was used:

$$P = \frac{3}{2} B_0 \left(\frac{V}{V_0}\right)^{-5/3} \left[1 - \left(\frac{V}{V_0}\right)^{-2/3}\right] \times \left\{\frac{3}{4}(B' - 4) \cdot \left[1 - \left(\frac{V}{V_0}\right)^{-2/3}\right] - 1\right\} \quad (1)$$

where $V_0$ is the unit cell volume at ambient pressure, $B_0$ is the bulk modulus, $B'$ is the pressure derivative of $B_0$. As a result of the approximation, the following parameters were obtained: $B_0$ = 37.0 ± 1.7 GPa, B' = 4 (fixed), $V_0/Z$ = 38.85 ± 0.15 Å$^3$, where $Z$ = 2 is the number of formula units in the crystal unit cell. The pressure at which the structural phase transition is $P_{tr}$ = 10.0 ± 0.5 GPa.

At higher pressures, the dependence $V(P)$ was approximated by the modified Birch-Murnaghan equation of state in the form:

$$P - 10 = \frac{3}{2} B_{10} \left(\frac{V}{V_{10}}\right)^{-5/3} \left[1 - \left(\frac{V}{V_{10}}\right)^{-2/3}\right] \times \left\{\frac{3}{4}(B' - 4) \cdot \left[1 - \left(\frac{V}{V_{10}}\right)^{-2/3}\right] - 1\right\} \quad (2)$$

where the subscript "10" indicates the parameter, values obtained above the structural transition at a pressure of 10 GPa. As a result of the approximation, the values of obtained parameters are: $B_{10}$ = 97.3 ± 3.8 GPa, $B'$ = 4 (fixed), $V_{10}/Z$ = 27.49 ± 0.04 Å$^3$, $Z$ = 4. The observed change in unit cell volume per formula unit is about 14.8 ± 0.2% (Fig. 9). For iron selenide FeSe in Ref. [8], the volume change during the structural transition at high pressure was about 11%.

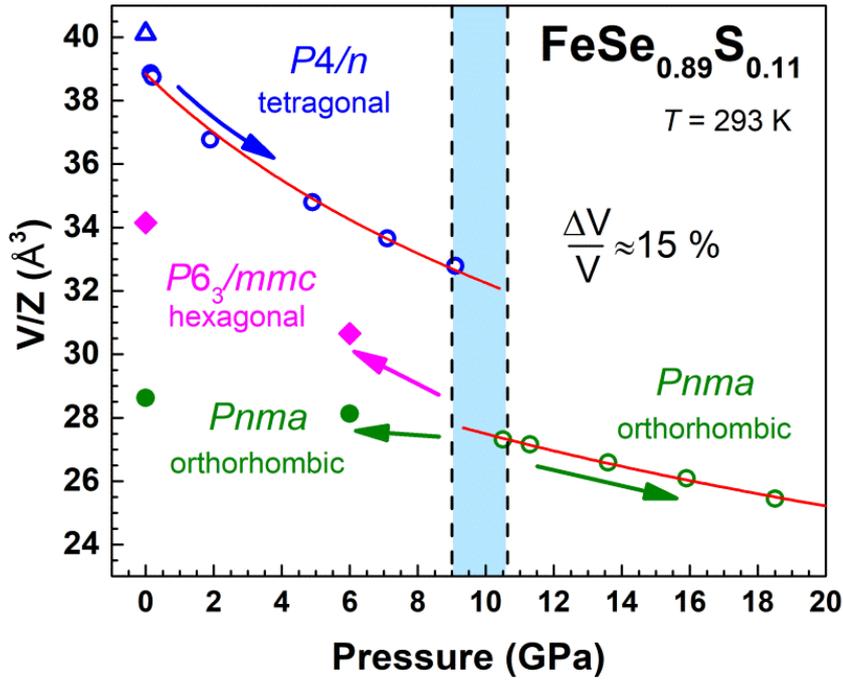

Fig.9. Dependence of the unit cell volume $V$ (per one formula unit) on pressure in a FeSe$_{0.89}$S$_{0.11}$ crystal at room temperature. For a tetragonal cell, the number of formula units is $Z = 2$. For an orthorhombic cell, it is $Z = 4$. The triangle symbol shows the value of $V$ obtained from a single-crystal experiment at ambient pressure. The solid lines show the approximation by the Birch-Murnaghan equation of state before and after the structural transition.

## 4. Discussion

Figure 10 shows the change in the crystal structures of polymorphic modifications of FeSe$_{0.89}$S$_{0.11}$ with increasing pressure in the range 0 - 18.5 GPa and subsequent decompression. At 10.5 GPa, a reconstructive phase transition with the rearrangement of the structure from tetragonal (Fig. 10a) to orthorhombic (Fig. 10b) occurs. At that, the coordination environment of Fe iron cations changes from tetrahedral to octahedral. With an increase in pressure in the range 10.5–18.5 GPa, the intermediate phase FeSe$_{0.89}$S$_{0.11}$ with a MnP-type structure is stable (Fig. 10b). This structure is an orthorhombic distortion of the hexagonal structure of the NiAs-type.

With a decrease in pressure, the FeSe$_{0.89}$S$_{0.11}$ sample recrystallizes, which is accompanied by a homogeneous deformation of the structure and a polymorphic transition of the distortion type from the orthorhombic structure of the MnP-type to a more symmetrical hexagonal structure of the NiAs-type (Fig. 10c). Probably, for the formation of the hexagonal phase, some exposure time of the sample at high pressure is required. Apparently, in our experiment during recording diffractograms in the pressure increase mode, a short exposure time (120 seconds) was not enough for the phase to fully form.

The formation of the hexagonal phase, partial conservation of the orthorhombic phase and the absence of the initial tetragonal phase upon removal of the applied pressure in the sample indicates a large hysteresis of the structural parameters during the phase transition in FeSe$_{0.89}$S$_{0.11}$. However, it is well established that there is no change of the $T_s$ anomaly under applied magnetic field either in transport and specific-heat measurements of FeSe systems, which might indicate that spin fluctuations are not involved directly in the structural transition. However, recent sound experimental studies on the origin of the nematic phase in iron chalcogenides reach opposing conclusions and this question remains highly debated [28, 32-34].

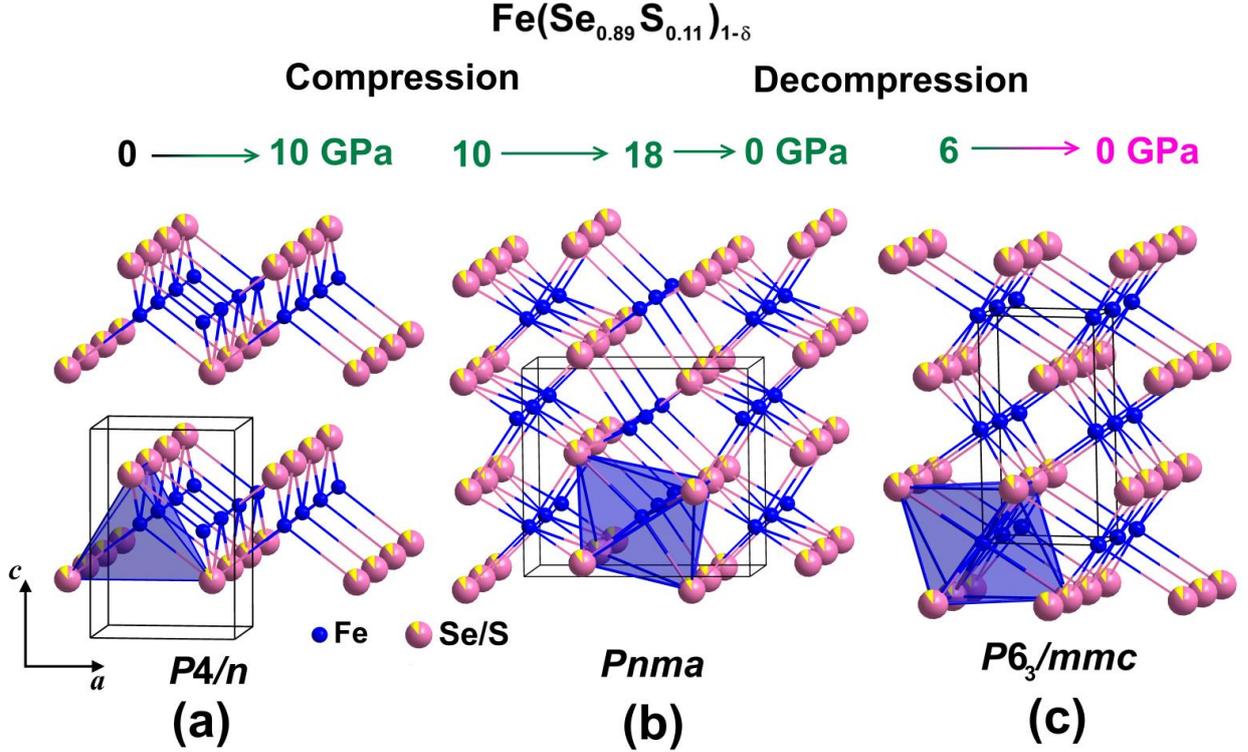

Fig. 10. Crystal structures of polymorphic FeSe$_{0.89}$S$_{0.11}$ modifications in the high-pressure region up to 18.5 GPa at room temperature: (a) the tetragonal modification structure (sp. gr. *P4/n*) is stable at pressures up to 10 GPa. (b) the orthorhombic structure (space group *Pnma*, the structural type MnP) is formed at a pressure above 10 GPa and remains up to 18.5 GPa, (c) the hexagonal structure (space group *P*6$_3$/*mmc*, the structural type NiAs) is a result of recrystallization of FeSe0.89S0.11 under a decrease in the applied pressure to 6 and 0 GPa.

## 5. Conclusion

The superconductor of iron selenide doped with sulfur FeSe$_{0.89}$S$_{0.11}$ with $T_C$ = 11 K was studied by X-ray diffraction under ambient conditions and at high pressures up to 18.5 GPa created in diamond anvil cells. It was found that, at ambient pressure and room temperature, FeSe$_{0.89}$S$_{0.11}$ has a tetragonal structure with the *P4/n* space group. According to heat capacity and electrical resistance data obtained at ambient pressure in FeSe$_{0.89}$S$_{0.11}$, a phase transition to the orthorhombic structure was detected at temperature of about $T_S \approx 71$ K. A similar transition in FeSe without sulfur was observed at $T_S \approx 80$ K. It was found that, with increasing pressure, the tetragonal modification FeSe$_{0.89}$S$_{0.11}$ (sp. gr. *P4/n*) is stable up to 10 GPa. It should be noted that even at a pressure of 0.15 GPa, a single-crystal sample is transformed into a polycrystal one, despite the use of helium as the mildest pressure-transmitting medium. At a pressure of 10.5 GPa, a reconstructive phase transition to an orthorhombic modification with a MnP-type structure (sp. gr. *Pnma*) occurs in FeSe$_{0.89}$S$_{0.11}$; however, the structural quality of the sample noticeably deteriorates due to its partial amorphization under pressure. With a successive decrease in the applied pressure to 6 GPa and 0 GPa in most of the FeSe$_{0.89}$S$_{0.11}$ sample, recrystallization occurs with the transition of the orthorhombic structure to the hexagonal one with the NiAs structural type (sp. gr. *P*6$_3$/*mmc*). In this case, FeSe$_{0.89}$S$_{0.11}$ does not return to the initial tetragonal modification with a decrease in pressure to ambient. Note that in the undoped iron selenide with a tetragonal structure, the hexagonal phase appears at pressures above 10 GPa [6]. However, in our experiment with the initially tetragonal FeSe$_{0.89}$S$_{0.11}$, the hexagonal phase occurs from the orthorhombic one when the pressure decreases from 18 GPa to ambient. The irreversibility of the phase transition in iron selenide doped with sulfur FeSe$_{0.89}$S$_{0.11}$, as well as the formation of the hexagonal phase after decompression, demonstrate a significant hysteresis of the structural parameters in this compound during the phase transition under pressure.

**Acknowledgments**

This work was supported by the Russian Science Foundation [grant number 17-72-20200] in the part of the sample preparation and the assembling of diamond anvil cells and by the Russian Ministry of Science and Higher Education within the State assignment of the FSRC "Crystallography and Photonics" of RAS in the part of the x-ray diffraction experiments. The authors acknowledge the beamline staff at the ESRF: Dr. M. Mezouar (ID27 beamline) for the assistance of the use of the beamline's equipment during high-pressure x-ray diffraction experiments and Dr. D. Chernyshov, Dr. Iu. Dovgaliuk (SNBL ID-01 beamline) for the assistance with single-crystal x-ray diffraction experiment. MAH acknowledge the financial support from the Swedish Research Council (VR) under the project number 2018-05393.